\documentclass[12pt,preprint]{aastex}
\usepackage[]{natbib,lscape,graphicx,graphics}

\usepackage[]{verbatim}

\shorttitle{Li in Metal-Poor Stars}
\shortauthors{Schaeuble \& King}

\begin{document}

\title{New Lithium Measurements in Metal-Poor Stars\altaffilmark{{\dagger}}} 
\author{Marc Schaeuble \& Jeremy R. King}
\affil{Department of Physics and Astronomy, Clemson University,
    Clemson, SC 29630-0978}
\email{marcs944@gmail.com,jking2@clemson.edu}
\altaffiltext{{$\dagger$}}{This paper includes data taken at The McDonald Observatory of The University of Texas at Austin.}

\begin{abstract}

We provide ${\lambda}6708$ \ion{Li}{1} measurements in 37 metal-poor stars, most of which 
are poorly-studied or have no previous measurements, from high-resolution and high-S/N 
spectroscopy obtained with the McDonald Observatory 2.1m and 2.7m telescopes. The typical line strength 
and abundance uncertainties, confirmed by the thinness of the Spite plateau manifested by our data and by comparison with previous measurements, are ${\le}4$ m{\AA} and ${\le}0.07-0.10$ dex respectively. 
Two rare moderately metal-poor solar-$T_{\rm eff}$ dwarfs, HIP 36491 and 40613, with significantly depleted but still detectable 
Li are identified; future light element determinations in the more heavily depeleted HIP 40613 may provide
constraints on the Li depletion mechanism acting in this star.  We note two moderately metal-poor and slightly 
evolved stars, HIP 105888 and G265-39, that appear to be analogs of the low-Li moderately metal-poor subgiant 
HD 201889.  Preliminary abundance analysis of G 265-39 finds no abnormalities that suggest the low Li content
is associated with AGB mass-transfer or deep mixing and $p$-capture.  We also detect line doubling in HIP 4754, 
heretofore classified as SB1.  

\end{abstract}

\keywords{Stars}

\slugcomment{Accepted for publication in The Publications of the Astronomical Society of the Pacific, volume 912 (February 2012)}

\section{INTRODUCTION}

The discovery of near-constant Li abundances in warm little-evolved metal-poor stars by \citet{SS82} quickly 
spawned a vigorous cottage industry seeking to derive the cosmological baryonic density under the assumption 
that the Li abundances in such stars reflect the product of Big Bang nucleosynthesis.  While WMAP observations 
\citep{Dunk2009}  have obviated the use of light element abundances to provide precision cosmological parameters, 
the study of metal-poor stellar Li abundances still has considerable import.  Even recent studies with warmer 
metal-poor $T_{\rm eff}$ scales, which result in higher Li abundances {\it ceteris paribus}, find that stellar estimates of the
primordial Li abundance is a factor of 3 smaller than that suggested by WMAP-based 
baryonic densities \citep{Hosford2010}.   The adequacy of our understanding of Big Bang nucleosynthesis remains 
unclear given this discrepancy.  

A salient question in considering this discrepancy is the degree to which metal-poor stellar Li abundances are 
post-primordial--i.e., possibly afflicted by stellar depletion, Galactic enrichment, or both.  Indeed, evidence exists 
that the Li abundances in the most metal-poor warm little-evolved stars is not constant, but declines with declining 
[Fe/H] \citep{Sbordone2010}.  While standard stellar models predict little, if any, stellar Li depletion in such 
stars, observational evidence suggests that such models suffer deficiencies in explaining Li abundances in cool
metal-poor dwarfs and evolved metal-poor stars \citep{Pilach1993,RD95}.  

Determination of Li abundances in large samples of metal-poor stars of various evolutionary state, mass, and 
metallicity are needed to provide the context in which to understand the relation of those abundances 
with the primordial value.  Large samples also yield rare examples of stars with anomalous light element abundances 
that can provide constraints on Galactic enrichment and stellar depletion processes \citep{K96,RD95,Boes2007,Koch2011}.  
Here, we make a modest contribution to expanding the size of such samples by providing Li measurements based on 
high-resolution spectroscopy in 37 metal-poor stars.  Our stars were selected from our existing observations 
connected with unrelated programs, and are metal-poor objects from the surveys of \citet{Carney1994} and 
\citet{Ryan1991} that have kinematics hotter than evinced by thin disk stars.  The stars selected are also 
poorly-studied with respect to light element abundances; the majority have no previous Li measurements.   

\section{OBSERVATIONAL DATA}

Spectra were obtained during observing runs in January and August 1994 and October 2004 with the 
McDonald Observatory 2.7-m Harlan J. Smith Telescope and its 2dcoude spectrograph \citep{Tull1995}, 
and in February and September 1994 with the McDonald Observatory 2.1-m Otto Struve Telescope and its 
Sandiford Cassegrain Echelle Spectrograph \citep{McCarthy1993}.  The nominal spectral resolution is 
$R{\sim}60,000$, except for the January 1994 2.7-m observations for which smaller slits and pre-Textronix 
CCDs with smaller pixels yielded resolutions of 67,000-90,000.  The per pixel S/N in the ${\lambda}6708$ 
\ion{Li}{1} region ranges from 80-450, and is in the smaller range 150-200 for most stars.  The 
{\sf echelle} package within IRAF\footnote{IRAF is distributed by the National Optical Astronomy 
Observatories, which are operated by the Association of Universities for Research in Astronomy, Inc., 
under cooperative agreement with the National Science Foundation.} was used to perform standard reductions 
including bias subtraction, flat fielding, order tracing and summation, and wavelength calibration to 
Th-Ar lamp spectra.  Examples of the spectroscopic data can be seen in Figure 1.  The stars considered
here are listed in Table 1; select cross-identifications are listed in the final column.
\marginpar{Fig1}
\marginpar{Tab1}

\section{ABUNDANCE ANALYSIS AND UNCERTAINTIES}

\subsection{General Approach}

While most metal-poor star Li abundance studies are conducted using measured line strengths, our
preference is to fit realistic models of the data (i.e., synthetic spectra) to the data and report
the equivalent widths associated with these model fits.  Synthetic spectra were calculated in LTE 
using an updated version of {\sf MOOG} \citep{Sneden1973} and the ${\lambda}6708$ \ion{Li}{1} region 
linelist from \citet{King1997} modified using more recent laboratory-based atomic data from VALD 
\citep{Kupka1999}, semi-empirical atomic data from Kurucz\footnote{http://kurucz.harvard.edu/linelists.html}, 
and the molecular CN data of \citet{Mandell2004}. We utilized model atmospheres interpolated within ATLAS9 
grids\footnote{http://kurucz.harvard.edu/grids.html} that correspond to the stellar parameters
described next. 

\subsection{Stellar Parameters}

Inasmuch as our main goal is fitting model spectra to the data in order to determine a line strength 
(rather than an abundance per se) that reflects the absorption flux, accurate stellar parameters are 
of little consequence.  In principle, the parameters could have a second order influence due to curve 
of growth effects and the presence of blending (primarily CN for the $\lambda$6708 \ion{Li}{1} feature 
at our resolution and given the small macroscopic line broadening associated with our stars).  However, 
even these second order effects are negligible: {\ }the weakness of the \ion{Li}{1} feature and the 
distribution of its absorbed flux over multiple hyperfine components nulls the curve of growth effects, 
and the metal-poverty of our stars nulls blending effects (especially so for diatomic CN blends).  While 
adopted stellar parameters need only be reasonably plausible for the purpose of measuring line strengths 
via model profile fits, it should be said that accurate stellar parameters are certainly required to 
determine reliable abundances from the syntheses.  These parameter-dependent abundances enable us to  
empirically examine our claimed estimated uncertainties by examining abundance scatter, compare our 
results with those of others, and identify any stars with anomalous Li abundances that may provide 
additional insight into metal-poor stellar Li depletion or Galactic or stellar Li production. 

The parameters were estimated through extant literature data as described below.  Even for the 
poorly-studied stars, color-based $T_{\rm eff}$ estimates on modern IRFM-based scales are easily determined, 
metallicity estimates from low-resolution spectroscopic measurements are available, and gravities
can be determined using isochrones and information about evolutionary status from HIPPARCOS parallaxes
and/or Stromgren photometry.  The proof of the relative quality of the parameters (most importantly $T_{\rm eff}$ 
for the Li abundance determinations) will be found in the pudding of the abundance results shown in Figure 3, 
which evinces only small scatter (${\pm}0.07$ dex) about the visible Spite plateau at warm $T_{\rm eff}$.  

\subsubsection{Effective Temperature}
Effective temperatures were estimated using four primary sources: excitation balance-based results 
from spectroscopic fine analyses in the literature (unavailable for many of our objects), Balmer line 
fitting-based results in the literature, the color-based values from \citet{Carney1994}, \citet{Alonso1996}, 
and \citet{Casagrande2010}. For a few stars, original color-based estimates were made using the new
new color-$T_{\rm eff}$ calibrations of \citet{Casagrande2010}; in those cases, appropriate reddening 
corrections were applied using the appropriate transformations from \citet{Schlegel1998} and extinctions 
or reddenings taken from \citet{Schuster2006}, \citet{SN89}, and/or \citet{Carney1994}.  Mean $T_{\rm eff}$ 
values, uncertainties in those means, and corresponding references are given in columns 2 and 3 of Table 1.  
Uncertainties based on one source are taken from the source, while those based on 2 sources are simply the 
difference between the source values. 

\subsubsection{Gravity}
Gravities were estimated from high-resolution spectroscopic fine analyses and/or Yale-Yonsei
isochrones \citep{Demarque2004} based on our $T_{\rm eff}$ value and/or $M_V$ inferred from HIPPARCOS 
parallax measurements or Stromgren calibrations.  In the former case,  log $g$ is 
in general double-valued depending on whether the star is on the main-sequence or subgiant branch.  
Such ambiguities were cleanly resolved based on the HIPPARCOS- and Stromgren-based $M_V$ values with
only a few exceptions: 

{\it G16-20}:  The spectroscopic gravity derived by \cite{Nissen2010} and that 
adopted on the basis of catalog-based distance and isochrone-based mass by \citet{Reddy2008} differ 
by a factor of 20, making it unclear whether this star is a dwarf or warm subgiant.  The Stromgren 
photometry of \citet{Schuster1993} suggests the star is a subgiant given its placement in the $c_1$ 
versus $(b-y)$ classification diagram of \citet{Schuster2004}. However, the extensive efforts of 
\citet{Arnadottir2010} to identify Stromgren-based evolutionary discriminants suggest that 
distinguishing between dwarfs and subgiants at $(b-y){\le}0.55$ (which holds for G 16-20) on the 
basis of Stromgren photometry cannot be done reliably.  We assume subgiant status.  

{\it G130-65}:  The log $g$ values inferred from $T_{\rm eff}$ and the \citet{Carney1994} distance 
(which assumes dwarf status) are in considerably better agreement than the subgiant-like log $g$ values 
inferred from the $T_{\rm eff}$ and the \citet{Schuster2006} Stromgren-based $M_V$.  We assume dwarf status.  

{\it G265-39}:  Given the negligible HIPPARCOS parallax, we assume subgiant status in 
estimating the gravity of this star from isochrones and our adopted $T_{\rm eff}$ value.  The so-derived log $g$ 
value also yields good agreement between Fe abundances derived from \ion{Fe}{1} and \ion{Fe}{2} 
lines; moreover, abundances of Ba and Y derived from gravity-sensitive singly ionized lines are in good
agreement with similarly metal-poor stars analyzed in \citet{Edvard1993}.  Details can be found in \S 4.  

Notes concerning the gravity estimates of a few other stars are made in Table 1.  In those cases, we believe
that dwarf/subgiant status is clearly resolved.  The gravity estimates and associated references/notes
can be found in columns 4 and 5 of Table 1.  Those stars with multiple log $g$ estimates, from both
spectroscopic fine analyses and parallaxes/isochrones, suggest that the log $g$ estimates are 
good to within $0.10-0.15$ dex; regardless, the Li line strengths and abundances are insensitive to the
adopted log $g$ values.  The 15 stars with Hipparcos parallaxes ${\ge}2$ their parallax uncertainties are 
plotted in the H-R diagram of Figure 2 with a selection of Yale-Yonsei isochrones \citep{Demarque2004} for context. 
\marginpar{Fig2}

\subsubsection{Metallicity and Microturbulence}

Metallicity estimates from the literature and their references are given in columns 6 and 7 of Table 1.  
Mean uncertainties are determined as they were for the $T_{\rm eff}$ values.  For metallicities coming from the 
\citet{Carney1994}estimates alone, the uncertainties are internal values based on their measurements 
from multiple spectra.  For stars with measurements from multiple independent sources, comparison of 
metallicity estimates suggests $0.15-0.2$ dex as a more realistic mean uncertainty for the \citet{Carney1994} 
values alone.  Microturbulence has negligible influence on the derived abundances and inferred line 
strengths.  In the interest of full disclosure, columns 8 and 9 list the values we adopted and their origin.  

\subsection{Results and Uncertainties}

Our object spectra are narrow-lined, showing no discernible signatures of rotation-dominated profiles.  
The synthetic spectra were smoothed to account for instrumental and thermal (and any small rotational) 
broadening by convolution with a Gaussian.  The FWHM used in the smoothing 
was empirically determined for each star by direct measurement of weak metal lines throughout the 
spectra, fits to the ${\lambda}6717$ \ion{Ca}{1} feature, and fits to the metal lines in the 
${\lambda}6104$ \ion{Li}{1} region using the linelist of \citet{KSHP10}.  Examples of the synthetic fits 
to the spectra can be seen in Figure 1.  The resulting LTE Li abundances are listed in column 10 of Table 1. 
The equivalent widths corresponding to the best fit synthetic profiles, as determined from traditional 
residual minimization compared to the observed data over the line profile, are given in column 11 of Table 1. 
NLTE abundance corrections (which do not impact the line strength measurements) were taken from 
\citet{Carlsson1994} and applied to the LTE abundances; the resulting NLTE abundances are listed in column 
12 of Table 1.  

Uncertainties in the LTE Li abundances are dominated by uncertainties in the smoothing, continuum 
placement, fitting, and $T_{\rm eff}$ estimates. Uncertainties in metallicity (including the direct effect
of blending), gravities, and microturbulence are negligible for the ${\lambda}6708$ \ion{Li}{1} line
in our stars.  The effects of uncertainties in our measurements of smoothing, our choice of continuum 
normalization, and our $T_{\rm eff}$ estimate (given in column 2 of Table 1) on the derived abundances
were measured by refitting syntheses accordingly adjusted; the abundance uncertainties
are attached to the LTE abundances in Table 1.  Sources of uncertainty in the line strength are those
listed above excluding those for $T_{\rm eff}$; these amount to a 10-15\% uncertainty in the reported 
line strengths.

Notes are made concerning two of the objects in our sample:\\
{\it HIP4754}: \citet{Latham2002} classify HIP 4754 as a single-lined spectroscopic binary with a 
347 day period. In our spectrum, the lines consistently evince a very weak blue asymmetry that can 
be seen in the ${\lambda}6708$ \ion{Li}{1} feature in Figure 1.  This asymmetry is confirmed in Fourier 
space: the cross-correlation function (CCF) formed from HIP 4754 and other warm stars in our sample 
(utilized as templates) shows a weak but distinct blue hump.  Fitting the CCFs with a two-component 
Gaussian indicates the radial velocity separation in our spectrum is ${\sim}5$ km s$^{-1}$.  

{\it G130-65}:  The red wing of the \ion{Li}{1} feature appears to exhibit an absorption asymmetry 
not seen in other lines (see Figure 1). Our fit to the profile ignores this asymmetry.  Additional spectroscopy would 
be desirable to confirm the presence or absence of a real asymmetry and any association with $^6$Li.   
We note that the asymmetry in the profile of HIP 36491 seen in Figure 1 is simply the $^7$Li hyperfine 
structure, which is manifest for this object due to the higher spectral resolution used when observing this 
star.

\section{DISCUSSION}

The main product provided here is the line strengths contained in column 11 of Table 1, which can 
be utilized in future homogenized meta-analyses of Li in metal-poor stars.  A few remarks
concerning the results can be made with the help of Figure 3, which plots the 
derived abundances versus $T_{\rm eff}$.  Metal-poor ([Fe/H]${\le}-1.29$) dwarfs or mildly-evolved subgiants 
are shown as solid squares; more metal-rich ($-0.92{\le}{\rm [Fe/H]}{\le}-0.44$) dwarfs or
little-evolved subgiants are shown as open squares.  Cooler, more highly-evolved subgiants 
and giants ($T_{\rm eff}{\le}5576$ K, log $g{\le}3.77$) are shown as open circles.  
\marginpar{Fig3}

\subsection{The Spite Plateau and Quality Estimates}

The metal-poor dwarfs evince the well-known pattern of near constant Li abundance for 
$T_{\rm eff}{\ge}5700$ K, referred to as the ``Spite plateau'' after \citet{SS82}, and declining 
Li abundance at cooler $T_{\rm eff}$ due to standard stellar model Li burning during the 
pre-main-sequence and main-sequence with the proportion of each being $T_{\rm eff}$- (or, 
more accurately, mass) dependent \citep{Deli90}, and the probable effects of rotationally-induced 
mixing \citep{RD95}. The scatter in Li on the metal-poor Spite plateau provides information about the 
quality of our results.  We confine our attention to the metal-poor dwarfs with $T_{\rm eff}{\ge}5692$ K
in order to avoid the effects of the most significant depletion.  We fit these data with a 2nd order 
polynomial that is shown as the dashed line in Figure 1.  The data exhibit a scatter of only 0.070 dex 
around this fit.  The larger expected value of 0.099 dex, based on the uncertainties in Table 1, suggests 
we may have overestimated the latter. Inasmuch as we believe the $T_{\rm eff}$ and fitting uncertainties have 
been realistically estimated, this indicates the subjectively assessed uncertainties in continuum normalization 
and smoothing are overestimated.  The uncertainties in $T_{\rm eff}$ in Table 1 lead to an estimate of an 
expected abundance scatter of 0.062 dex.  Subtracting this in quadrature from the observed scatter implies 
the line strengths uncertainties (which are controlled by uncertainties in fitting, continuum normalization, 
and smoothing) are ${\le}8$\%-- translating to an equivalent width uncertainty of ${\le}$2-4 m{\AA}.  

An alternative estimate of the line strength uncertainties is provided by comparison of our measurements with 
the eleven previous measurements in the notes to Table 1.  The standard deviation of the differences between 
our line strength differences and those of others is ${\pm}5.4$ m{\AA} (the average difference, in the sense
of our measures minus others', is $-2.6$ m{\AA}).  Assuming equivalent uncertainties 
in our measurements and others' implies a typical uncertainty in our measurements of ${\pm}3.8$ m{\AA}--in 
good accord with the Spite plateau-based estimate.  

Our mean Spite plateau abundance of log $N$(Li)${\sim}2.3$ agrees with numerous other measures of similarly 
warm and metal-poor stars.  We do not address here the pregnant issue of the inconsistency of such a 
primordial Li abundance inferred from stellar measurements with that implied by WMAP observations for 
two reasons.  First, the absolute $T_{\rm eff}$ scale of metal-poor stars remains uncertain, perhaps by 
as much as 100-200 K \citep{King93,Casagrande2010,Hosford2010}, especially for stars with [Fe/H]${\le}-2$; 
this scale has significant implications for the primordial Li abundance derived from metal-poor stars 
\citep{King94,Melendez2004}.  Second, the Li abundance in the warmest little-evolved halo stars may decline 
with decreasing metallicity at ultra-low metallicites-- behavior predicted by \citet{Ryan1999}, observed 
by \citet{Sbordone2010}, but disputed by \citet{Melendez2010}.  Whether the Spite plateau is in fact a plateau 
at very low [Fe/H] remains uncertain.  However, we simply note that the slope of our Spite plateau stars' 
NLTE Li abundances with [Fe/H], 0.14${\pm}0.05$ dex/dex, is in good agreement with the value ($0.15{\pm}0.05$) 
derived by \citet{Hosford2009} for main-sequence stars in the metallicity range 
$-3.3{\le}{\rm [Fe/H]}{\le}-2.3$ using temperatures derived from Fe excitation under the LTE assumption, and 
with the value ($0.14{\pm}0.12$) derived by \citet{Hosford2010} using temperatures derived from Fe excitation 
in a NLTE framework ignoring H collisions.

\subsection{The Warm Moderately Metal-Poor Dwarfs}

The four more moderately metal-poor ($-0.92{\le}{\rm [Fe/H]}{\le}-0.44$) dwarfs evince a real 
factor of ${\sim}7$ scatter in their Li abundance.   From a theoretical perspective, such a spread
can naturally be accommodated by the inclusion of rotationally-induced mixing (with or without an 
age spread) in stellar models.  Figures 4 and 5 of \citet{Pin92} show that such models produce a 
wider dispersion in pre-main-sequence and main-sequence depletion compared to more metal-poor 
models.  From an observational context, the situation is more complex and interesting.  Nearly all 
field dwarfs with near-solar $T_{\rm eff}$ values in the metallicity range above have Li abundances 
of ${\sim}2.1$ with modest scatter (${\sim}0.2$ dex), as seen in Figure 2 of \citet{LHE91}.  Stars with 
significantly lower abundances (like HIP 36491 and 40613 in our sample) appear to be very rare, as seen 
in Figure 2 of \citet{LHE91}.  

What is interesting about our
measurements of HIP 36491 and 40613 is that they reveal Li to be more heavily depleted in these stars
but still detectable.  As described in detail in \S 5.1 of \citet{Stephens97}, observations of Be in such 
stars can place constraints on numerous candidate mechanisms responsible for the Li depletion in 
these objects.  The Be abundance derived by \citet{Smil2009} places HIP 36491 just above the mean Galactic 
Be-Fe trend at our metallicity in both their Figure 10 as well as in Figure 2 of the independent study of
\citet{Boes2010}.  Any Be depletion in this star is apparently limited to ${\le}0.1$ dex if one compares
the abundance to the upper envelope of the \citet{Smil2009} and \citet{Boes2010} Be-Fe data.  The 
inequality of inferred Li and Be depletion, ${\sim}0.4$ dex versus ${\sim}0$ dex, would seem to exclude 
diffusion as a causal mechanism.  The relative depletions are qualitatively consistent with 
mass loss, gravity waves, meridional circulation, and rotationally-induced mixing.  Determination of 
the Be abundance in the more highly Li-depleted HIP 40613 would be of interest in perhaps providing
stronger constraints on these remaining depletion mechanisms.  

\subsection{The Evolved Stars}

The Li abundances of the 5 subgiants and giants with $T_{\rm eff}{\le}5600$ K are lower than the Spite plateau 
values, and consistent with the values expected from non-diffusive dilution (e.g. Figure 8 of Pilachowski, 
Sneden \& Booth 1993) that occurs when the stellar surface convection zone dips into deeper hotter Li-depleted 
regions. The two more metal-rich ([Fe/H]${\sim}-0.7$) subgiants with $T_{\rm eff}{\ge}5700$ K, HIP 105888
and G 265-39, show Li abundances significantly lower than the trend formed by the other subgiants and
the aforementioned dilution models.  These two stars are reminiscent of HD 201889, a subgiant with similar
$T_{\rm eff}$ and [Fe/H] that also demonstrates a rare anomalously low Li abundance compared with other warm
subgiants at this $T_{\rm eff}$ in Figure 7 of \citet{Pilach1993}. 

A possible explanation for the low Li abundances in HIP 105888 and G 265-39 is an environmental origin--
perhaps these stars were contaminated with the Li-depleted products of a former AGB companion or have undergone 
unexpected $p$-capture processing and deep mixing that have destroyed Li and brought this Li depleted material to the surface.  The first possibility might lead to $s$-process enhancements, while the second possibility might also be 
accompanied by Ne$\rightarrow$Na and O$\rightarrow$N cycling.  We have conducted a 
preliminary search for these signatures in G 265-39, but failed to find them.  

We measured the line strengths of a few \ion{Fe}{1}, \ion{Fe}{2}, \ion{O}{1}, \ion{Na}{1}, \ion{Zr}{2}, \ion{Y}{2}, and 
\ion{Ba}{2} lines in G 265-39 and a daytime sky (solar proxy) spectrum acquired with the McDonald 2.7-m.  The 
lines, line strengths, and resulting absolute logarithmic number abundances derived from the line strengths using
{\sf MOOG} are listed in Table 2.  Atomic data is taken from \citet{Edvard1993}, \citet{King1998}, and \citet{Sch06}.  
The mean Fe abundance from the 5 lines is [Fe/H]$=-0.73$, which is in outstanding agreement with the value adopted in 
Table 1 ($-0.71$) within the mean measurement uncertainty (${\pm}0.05$ dex) alone.  Because of the particular sensitivities 
of the individual lines, the complementary ratios [O I,Zr II, Y II, Ba II/Fe II] and [Na I/Fe I] are essentially free of
uncertainties in the stellar parameters (at least compared to measurement uncertainties).  The ratios\footnote{Our
permitted \ion{O}{1}-based ratio was placed on ${\lambda}6300$ forbidden \ion{O}{1}-based scale of 
\citet{Edvard1993} using their equation 11.} in G 265-39 are similar to those exhibited by stars 
of the same [Fe/H] in Figure 15 of \citet{Edvard1993}.  In particular, O does not appear underabundant 
nor does Na appear overabundant as might occur if our star was contaminated by material having undergone 
O${\rightarrow}$N and Ne${\rightarrow}$Na cycling and deep mixing (either {\it in situ} or from a former 
AGB companion); nor is there any indication of an $n$-capture element overabundance that might accompany 
contamination by material from a former AGB companion.  

More extensive surveys of metal-poor stars are needed to identify larger numbers of objects with anomalous Li 
abundances.  Subsequent or accompanying determination of abundances of a suite of elements in these stars is needed to 
understand what information they provide about the effects of stellar physics and Galactic chemical evolution 
on Li abundances.  Coupled with a solid theoretical framework, such data will be required to establish 
the existence or not of a decline of stellar Li at extremely low metallicities and to place the current apparent 
mismatch of these Li abundances with WMAP results into an appropriate cosmological context.  

\acknowledgments
MS and JRK gratefully acknowledge support for this work from NSF grant AST 09-08342 to JRK.  The 
observations were originally supported by NASA through the grant HF-1046.01-93A to JRK from the Space 
Telescope Science Institute, which is operated by the Association of Universities for Research in Astronomy, 
Inc.~under NASA Contract No. NAS 5-26555.

%Figure 1
\begin{figure}
\includegraphics[width=6.5in]{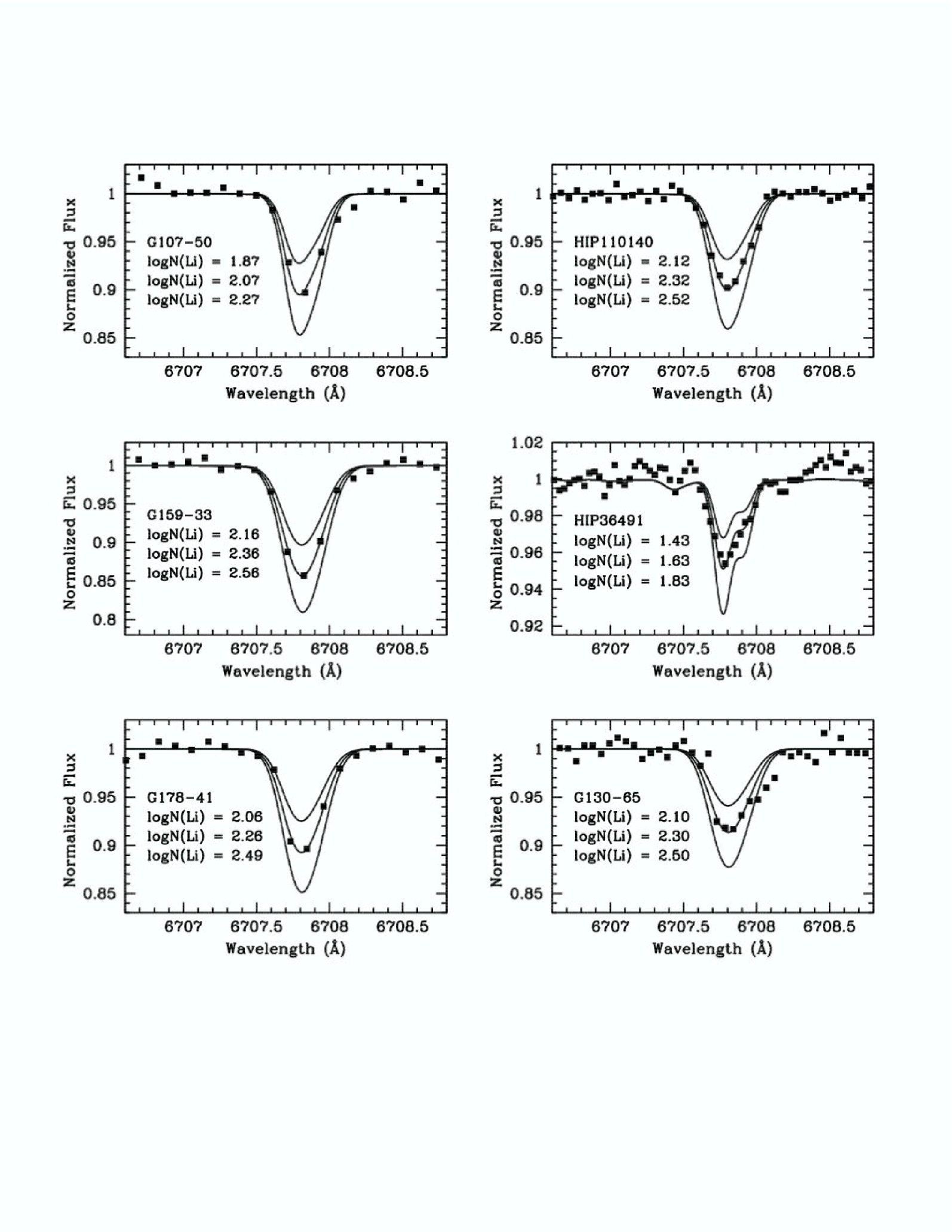}
\caption{Sample ${\lambda}6708$ \ion{Li}{1} region data (solid points) and syntheses (lines) with input Li abundances stepped
by 0.2 dex.}
\label{figSyns}
\end{figure}

%Figure 2
\begin{figure}
\includegraphics[width=6.5in]{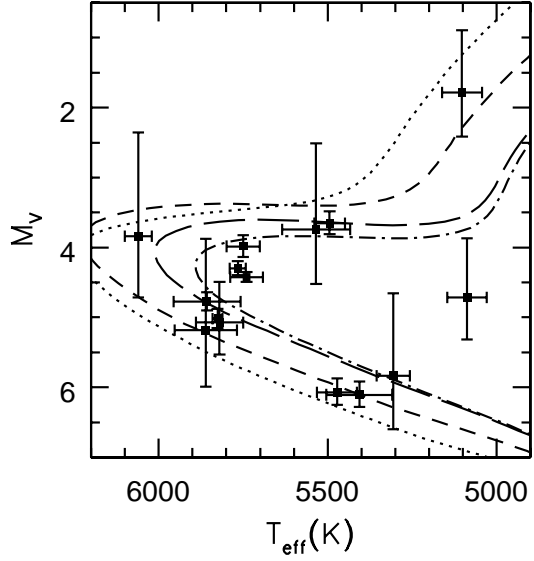}
\caption{Our sample tars having Hipparcos parallaxes satisfiying ${\pi}/{\sigma}_{\pi}{\ }{\ge}2$ in an 
H-R diagram using Hipparcos-based absolute magnitude and our effective temperatures; error bars reflect 
1${\sigma}$ level uncertainties in the parallaxes and $T_{\rm eff}$ values.  Several Yale-Yonsei 
[${\alpha}$/Fe]$=+0.3$ isochrones are shown for reference: [Fe/H]$=-1.5$, 12 Gyr (short dashed line); 
[Fe/H]$=-0.9$, 8 Gyr (medium dashed line); [Fe/H]$=-0.5$, 8 Gyr (long dashed line); and [Fe/H]$=-0.5$, 
10 Gyr (dot-dashed lined).}
\label{figHR}
\end{figure}

%Figure 3
\begin{figure}
\includegraphics[width=6.5in]{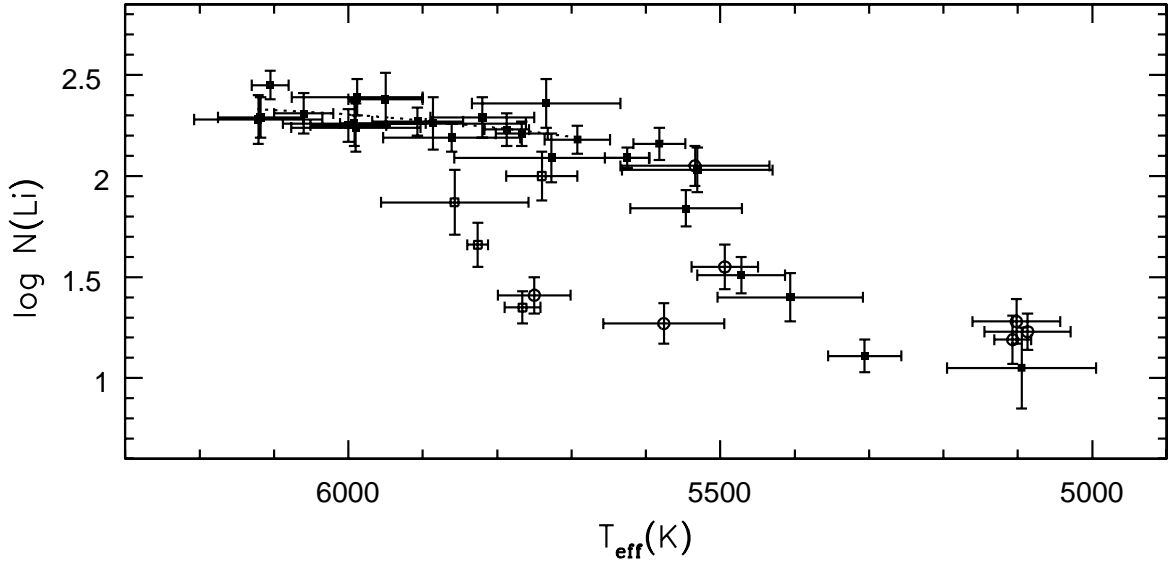}
\caption{Our synthesis-based NLTE Li abundances are plotted versus $T_{\rm eff}$.  Filled squares are metal-poor
([Fe/H]${\le}-1.26$) dwarfs or warm little-evolved subgiants.  Open squares indicate more metal-rich 
($-0.92{\le}{\rm [Fe/H]}{\le}-0.44$) dwarfs.  Open circles designate subgiants or giants with $T_{\rm eff}{\le}5600$ K.}
\label{figAbunds}
\end{figure}

%%Tables
%%Table 1
\begin{landscape}
\begin{deluxetable}{l c l c l c l c l c c c l}
\tablewidth{0 pt}
\tablecaption{Metal-Poor Sample and Parameters}
\tablehead{
\multicolumn{1}{l}{Name} &
\multicolumn{1}{c}{$T_{\rm eff}$} &
\multicolumn{1}{c}{Ref} &
\multicolumn{1}{c}{log $g$} &
\multicolumn{1}{c}{Ref}  &
\multicolumn{1}{c}{[Fe/H]} & 
\multicolumn{1}{c}{Ref} & 
\multicolumn{1}{c}{${\xi}$} & 
\multicolumn{1}{c}{Ref} & 
\multicolumn{1}{c}{$A$(Li)} & 
\multicolumn{1}{c}{EEW} & 
\multicolumn{1}{c}{$A$(Li)} & 
\multicolumn{1}{c}{Notes} \\
\multicolumn{1}{l}{} &
\multicolumn{1}{c}{K} &
\multicolumn{1}{c}{} &
\multicolumn{1}{c}{cgs} &
\multicolumn{1}{c}{} & 
\multicolumn{1}{c}{} & 
\multicolumn{1}{c}{} & 
\multicolumn{1}{c}{km/s} &
\multicolumn{1}{c}{} & 
\multicolumn{1}{c}{LTE} & 
\multicolumn{1}{c}{m{\AA}} & 
\multicolumn{1}{c}{NLTE} & 
\multicolumn{1}{c}{} \\
}
\startdata
G158-30  & $5546{\pm}75$ & 1 & 4.50 & 1 & $-1.29{\pm}0.07$ & 1 & $0.4$ & 1 & $1.79{\pm}0.09$ & 23.4 & 1.84 & 2; G31-26 \\
G130-65  & $6121{\pm}86$ & 3,4 & 4.50 & 5 & $-2.33{\pm}0.03$ & 3 & 1.4 & 6 & $2.30{\pm}0.12$ & 28.1 & 2.28 & 7 \\
HIP4754  & $5534{\pm}100$ & 8 & 3.50 & 9 & $-1.72{\pm}0.10$ & 8 & 1.5 & 6 & $1.99{\pm}0.10$ & 35.3 & 2.05 & 10; G33-30 \\ 
G2-38      & $5950{\pm}50$ & 11 & 4.57 & 11 & $-1.29{\pm}0.06$ & 11 & 1.3 & 6 & $2.39{\pm}0.13$ & 39.0 & 2.38 & 12; G33-60,G34-31 \\ 
G172-58  & $6118{\pm}57$ & 3,8,13 & 4.38 & 14 & $-1.57{\pm}0.25$ & 3,8,13 & 1.4 & 6 & $2.31{\pm}0.10$ & 28.4 & 2.29 & 15; BD$+$47{\ }435 \\
G133-45  & $5767{\pm}35$ & 3,16 & 4.57 & 5 & $-1.50{\pm}0.03$ & 3 & 1.2 & 6 & $2.20{\pm}0.06$ & 36.5 & 2.21 & \\
G159-33  & $5734{\pm}100$ & 8 & 4.63 & 5 & $-1.54{\pm}0.10$ & 8 & 1.2 & 6 & $2.36{\pm}0.12$ & 48.8 & 2.36 & G71-55 \\
HIP12710 & $6000{\pm}51$ & 11,17,18,19 & 4.26 & 11,17,18,19 & $-1.83{\pm}0.13$ & 11,17,18,19 & 1.40 & 11,18,19 & $2.26{\pm}0.08$ & 30.0 & 2.25 & 20; G4-36,G76-25 \\
G36-47     & $5907{\pm}61$ & 3 & $4.55$ & 5 & $-1.63{\pm}0.14$ & 3,21 & 1.3 & 6 & $2.27{\pm}0.07$ & 33.8 & 2.27 & G 37-17 \\
G95-11     & $5531{\pm}101$ & 3 & 4.70 & 5 & $-2.09{\pm}0.05$ & 3 & 1.1 & 6 & $2.00{\pm}0.11$ & 35.4 & 2.03 & 7 \\
G37-37     & $5990{\pm}87$ & 1 & $3.76$ & 1 & $-2.36{\pm}0.06$ & 1 & 1.55 & 1 & $2.24{\pm}0.12$ & 30.0 & 2.24 & 22; G95-33 \\
G191-55  & $5787{\pm}30$ & 3,13,23 & 4.54 & 5,13,23 & $-1.74{\pm}0.12$ & 3,13,23 & 1.2 & 13,23 & $2.23{\pm}0.08$ & 37.1 & 2.23 & hh \\
G107-50  & $5727{\pm}131$ & 3 & 4.63 & 5 & $-2.25{\pm}0.03$ & 3 & 1.2 & 6 & $2.07{\pm}0.12$ & 30.6 & 2.09 & 7 \\
HIP36491    & $5826{\pm}14$ & 11,3,4,25 & 4.41 & 11,26,25 & $-0.92{\pm}0.03$ & 11,3,27,25,28,29,30 & 1.3 & 11,27,25,28 & $1.63{\pm}0.11$ & 10.8 & 1.66 & 31; G88-31, G89-25 \\
G234-24 & $5988{\pm}88$ & 3,4 & 4.47 & 5 & $-1.60{\pm}0.04$ & 3 & 1.3 & 6 & $2.41{\pm}0.09$ & 38.9 & 2.39 & \\
HIP40613    & $5766{\pm}24$ & 3,4,32,33,34 & 4.29 & 26,32,34,35 & $-0.58{\pm}0.02$ & 3,36,37,32,33,34,35 & 1.3 & 6 & $1.32{\pm}0.08$ & 6.9 & 1.35 & 38; G113-24 \\
HIP45069 & $6105{\pm}25$ & 39,40 & 3.87 & 39,40 & $-1.41{\pm}0.05$ & 39,40 & 1.40 & 39,40 & $2.47{\pm}0.07$ & 34.1 & 2.45 & G115-58 \\
HIP47316 & $5992{\pm}96$ & 3,4,16 & 4.47 & 6 & $-1.57{\pm}0.05$ & 3,41 & 1.3 & 6 & $2.27{\pm}0.11$ & 30.7 & 2.26 & G195-34 \\
HIP49371 & $5102{\pm}59$ & 11,4,25,32 & 2.66 & 11,25,32,42 & $-1.83{\pm}0.06$ & 11,25,32,43,44,45,42 & 1.6 & 4,25,32,45 & $1.13{\pm}0.11$ & 15.0 & 1.28 & HD87140 \\
HIP63100  & $5861{\pm}92$ & 3,4 & 4.57 & 5 & $-1.78{\pm}0.11$ & 3,46 & 1.3 & 6 & $2.19{\pm}0.07$ & 31.6 & 2.19 & G60-48,G61-20 \\
G150-50 & $5095{\pm}100$ & 3 & 4.77 &  5 & $-1.85{\pm}0.03$ & 3 & 1.0 & 6 & $0.95{\pm}0.20$ & 10.5 & 1.05 & 47 \\
G178-41 & $5886{\pm}125$ & 3,4 & 4.58 & 5 & $-2.29{\pm}0.16$ & 3,4,48 & 1.3 & b & $2.26{\pm}0.13$ & 35.3 & 2.26 & \\
HIP74588 & $5107{\pm}25$ & 23,39 & 3.21 & 23,39 & $-1.28{\pm}0.04$ & 23,39 & 1.3 & 23,39 & $1.05{\pm}0.12$ & 14.1 & 1.19 & 49; G179-22 \\
G16-20  & $5625{\pm}30$ & 50 & $3.64$ & 50 & $-1.42{\pm}0.04$ & 50 & 1.5 & 50 & $2.04{\pm}0.05$ & 34.9 & 2.09 & \\
G168-26 & $5582{\pm}35$ & 3 & 4.65 & 5 & $-1.80{\pm}0.04$ & 3 & 1.1 & 6 & $2.14{\pm}0.08$ & 42.3 & 2.16 & \\
HIP85757 & $5494{\pm}45$ & 3,4,25 & 3.77 & 25,26 & $-0.71{\pm}0.05$ & 3,36,27,25,29,37 & 1.5 & 6 & $1.48{\pm}0.11$ & 16.6 & 1.55 & G20-6, G19-7 \\
HIP97747 & $5087{\pm}58$ & 4,51,52 & 2.85 & 53,51,52,54 &  $-1.51{\pm}0.05$ & 23,51,52,54 & 1.5 & 6 &  $1.08{\pm}0.09$ & 14.8 & 1.23 & 55; G23-14,G92-33 \\
G265-39  & $5576{\pm}81$ & 3,4 & 3.77 & 26,56 & $-0.71{\pm}0.05$ & 3 & 1.5 & 6 & $1.21{\pm}0.10$ & 7.6 & 1.27 & HD200544 \\
HIP103269 & $5472{\pm}59$ & 11,3,4,25,57 &  4.67 & 11,5 & $-1.72{\pm}0.03$ & 11,3,21,58,27,25,57 & 1.1 & 11,27,25,57 & $1.45{\pm}0.09$ & 13.5 & 1.51 & 59; G212-7 \\
HIP105488 & $5820{\pm}70$ & 3,8 & 4.52 & 5 & $-1.48{\pm}0.14$ & 3,8,21 & 1.3 & 6 & $2.29{\pm}0.10$ & 39.0 & 2.29 & G187-40 \\
HIP105888 & $5750{\pm}49$ & 11,3,4,27,60,25,57 & 4.01 & 11,26 & $-0.74{\pm}0.05$  & 11,3,27,60,25,57,28,29 & 1.5 & 6 & $1.37{\pm}0.09$ & 7.5 & 1.41 & 61; G25-29,G93-9 \\
HIP106924 & $5406{\pm}98$ & 8,3,58,27,62 & 4.70 & 5,63 & $-1.84{\pm}0.08$ & 8,3,58,27,62 & 1.1 & 6 & $1.34{\pm}0.12$ & 12.0 & 1.40 & G231-52 \\
G214-5 & $5692{\pm}44$ & 3 & 4.60 & 5 & $-1.91{\pm}0.22$ & 3,21 & 1.2 & 6 & $2.17{\pm}0.07$ & 37.8 & 2.18 & \\
HIP110140 & $6060{\pm}40$ & 50,3,4,64 & 4.00 & 5 & $-1.50{\pm}0.04$ & 50,3,64,65,66 & 1.5 & 50,65 & $2.32{\pm}0.10$ & 30.5 & 2.31 & 67; G18-39 \\
HIP111332 &  $5857{\pm}99$ & 3,4 & 4.42 & 26 & $-0.44{\pm}0.04$ & 3,21 & 1.3 & 6 & $1.85{\pm}0.16$ & 16.8 & 1.87 & G156-19 \\
HIP116259 & $5740{\pm}48$ & 3,4 & 4.42 & 26 & $-0.55{\pm}0.03$ & 3 & 1.2 & 6 & $1.97{\pm}0.12$ & 25.9 & 2.00 & 68,69; G 171-3 \\
HIP117150 & $5306{\pm}49$ & 3 & 4.69 & 5 & $-1.62{\pm}0.05$ & 3 & 1.1 & 6 & $1.03{\pm}0.08$ & 7.9 & 1.11 & G130-7 \\
\enddata
\tablerefs{{\ }\\
(1) \citet{Stephens2002}\\
(2) $A$(Li)$=1.59$, EW(Li)$=16.5{\pm}1.0$ \citet{Boesgaard2005} \\
(3) \citet{Carney1994} \\
(4) \citet{Casagrande2010} \\
(5) Estimated from 12 Gyr [${\alpha}$/Fe]$=+0.3$ Yale-Yonsei isochrones \citep{Demarque2004} with \citet{Lejeune1998} color transformations using the $T_{\rm eff}$ and assuming evolutionary status of \citet{Carney1994} and/or 
$M_V$ value of \citet{Schuster2006} and/or the HIPPARCOS-based parallax \\
(6) adopted here \\
(7) The Yale-Yonsei dwarf-like log $g$ values inferred from $T_{\rm eff}$ and distance are in considerably better agreement than the subgiant-like log $g$ values inferred from the $T_{\rm eff}$ and \citet{Schuster2006} $M_V$ 
values.  We thus assume dwarf evolutionary status and adopt the \citet{Carney1994}-based values.\\
(8) \citet{Reddy2008}\\
(9) Estimated from 12 Gyr [${\alpha}$/Fe]$=+0.3$ Yale-Yonsei isochrones \citep{Demarque2004} with \citet{Lejeune1998} color transformations using the $T_{\rm eff}$ and assuming subgiant/dwarf status inferred by \citet{Reddy2008} \\
(10) SB 1 with $P=347$d,$e=0.38$ according to \citet{Latham2002}; the double-lined nature of the star is detected in our spectrum \\
(11)  \citet{Axer1994}\\
(12) Binary with 25 arcsec separation and 5.5 photographic magnitude brightness difference \citep{Allen2000}
(13) \citet{Zhang2005} \\
(14) The log $g$ value of \citet{Reddy2008} is 0.5 dex larger than that implied for their spectroscopic $T_{\rm eff}$ by the Yale-Yonsei isochrones.  The log $g$ value is determined from our $T_{\rm eff}$ value assuming dwarf status implied by the results of \citet{Reddy2008} and \citet{Zhang2005}. \\
(15) \citet{Zhang2003} find EW(Li)$=37.4$ m{\AA}, $A$(Li)$=2.20$ \\
(16) \citet{Alonso1996} \\
(17) \citet{Cenarro2007}\\
(18) \citet{Ivans2003}\\
(19) \citet{James2000}\\
(20) \citet{Ivans2003} find EW(Li)=32.7 m{\AA}, $A$(Li)$=2.35$ \\
(21) \citet{Schuster2006} \\
(22) \citet{Boesgaard2005} find $A(Li)=2.28$, EW(Li)$=34.5{\pm}1.4$ \\
(23) \citet{Simmerer2004} \\
(24) Binary with 0.8 arcsec separation and 2 photographic magnitude brightness difference resolved by speckle interferometry \citep{Rastegaev2007}
(25) \citet{Fulbright00} \\
(26) Gravity estimated from our adopted $T_{\rm eff}$ and/or HIPPARCOS-based $M_V$ value using an 8 Gyr Yale-Yonsei isochrone. \\
(27) \citet{Gratton2003} \\
(28) \citet{Zhang2006} \\
(29) \citet{Clem1999} \\
(30) \citet{Jonsell2005} \\
(31) \citet{Fulbright00} finds EW(Li)$=13.0$, $A$(Li)$=1.70$ \\
(32) \citet{Gratton1996} \\
(33) \citet{Bensby2005} \\
(34) \citet{Sousa2011} \\
(35) \citet{Edvard1993} \\
(36) \citet{Ramirez2007} \\
(37) \citet{Reddy2006} \\
(38) The $T_{\rm eff}$ and log $g$ values of \citet{Axer1994} are 200 K hotter and ${\ge}0.7$ dex smaller than other photometric and spectroscopic determinations; we do not utilize their values here.
(39) \citet{Zhang2009} \\
(40) \citet{Ishigaki2010} \\
(41) \citet{SN89} \\
(42) \citet{Tomkin92} \\ 
(43) \citet{Pilach1993} \\
(44) \citet{Cavallo97} \\
(45) \citet{Gratton2000} \\
(46) \citet{Beers1999} \\
(47) $T_{\rm eff}$ uncertainty is adopted \\
(48) \citet{Ryan1991} \\
(49) The HIPPARCOS parallax and spectroscopic gravity estimates indicate the dwarf status inferred from 
\citet{Carney1994} is incorrect. \\
(50) \citet{Nissen2010}\\
(51) \citet{Carney1997} \\
(52) \citet{Yong2003} \\
(53) Gravity estimated from both our adopted $T_{\rm eff}$ and/or the HIPPARCOS-based $M_V$ value using the 12 Gyr Yale-Yonsei isochrone. \\
(54) \citet{Jones1995} \\
(55) \citet{Carney1997} find EW(Li)$=11.2$ m{\AA}, $A$(Li)$=0.89{\pm}0.15$ \\
(56) The low Li and negligible parallax suggest subgiant status. \\
(57) \citet{Mishenina01} \\
(58) \citet{VF2005} \\
(59) \citet{Fulbright00} finds EW(Li)$=16.0$, $A$(Li)$=1.34$; \citet{Shi2007} find $A$(Li)$=1.60$ 
(60) \citet{Fuhrmann98} \\
(61) \citet{Fulbright00} finds EW(Li)$=7.9$ m{\AA}, $A$(Li$)=1.37$; \citep{Gutierrez1999} find EW(Li)$<13$ m{\AA} \\
(62) \citet{Mish2000} \\
(63) The spectroscopic gravity of \citet{VF2005} is a factor of 2.5 smaller than that implied by the Yale-Yonsei isochrones for their $T_{\rm eff}$ value.  We adopt the isochrone value implied by our adopted $T_{\rm eff}$ and the HIPPARCOS-based $M_V$ value. \\
(64) \citet{Nissen2007} \\
(65) \citet{Nissen2004} \\
(66) \citet{Caffau2005} \\
(67) \citet{RMB1988} fine EW(Li)$=37$ m{\AA}, $A(li)=2.24$ \\
(68) \citet{Latham2002} find SB1 with $P=6083$d,$e=0.52$; \citet{Horch2002} used speckle interferometry to detect a companion with ${\rho}=0.15$ arcsec separation. \\
(69) \citet{Gutierrez1999} find EW(Li)$=26$ m{\AA}; \citet{White2007} find EW(Li)$=37$ m{\AA} \\
}
\label{Metal-Poor Sample}
\end{deluxetable}
\end{landscape}

%Table2
\begin{deluxetable}{l c c c c c c c}
\tablewidth{0 pt}
\tablecaption{Abundances in G 265-39}
\tablehead{
\multicolumn{1}{l}{Species} &
\multicolumn{1}{c}{${\lambda}$} &
\multicolumn{1}{c}{${\chi}$} & 
\multicolumn{1}{c}{log $gf$} & 
\multicolumn{1}{c}{EW$_{\odot}$} & 
\multicolumn{1}{c}{log $N_{\odot}$} & 
\multicolumn{1}{c}{EW$_{\star}$} & 
\multicolumn{1}{c}{log $N_{\star}$} \\
\multicolumn{1}{c}{} & 
\multicolumn{1}{c}{{\AA}} &
\multicolumn{1}{c}{eV} &
\multicolumn{1}{c}{} &
\multicolumn{1}{c}{m{\AA}} &
\multicolumn{1}{c}{} &
\multicolumn{1}{c}{m{\AA}} &
\multicolumn{1}{c}{} \\
}
\startdata
\ion{O}{1} & 7771.95 & 9.15 & $+0.369$ & 74.0 & 8.92 & 53.7 & 8.57 \\
\ion{O}{1} & 7774.17 & 9.15 & $+0.223$ & 62.7 & 8.89 & 43.8 & 8.54 \\
\ion{O}{1} & 7775.39 & 9.15 & $+0.001$ & 52.6 & 8.93 & 35.9 & 8.59 \\
\ion{Na}{1} & 6154.22 & 2.10 & $-1.61$ & 39.8 & 6.36 & 19.5 & 5.88 \\
\ion{Na}{1} & 6160.75 & 2.10  & $-1.31$ & 60.0 & 6.34 & 33.1 & 5.87 \\
\ion{Fe}{1} & 5141.75 & 2.42 & $-2.19$ & 94.3 & 7.59  & 69.7 & 6.79 \\
\ion{Fe}{1} & 6151.62 & 2.18 & $-3.33$ & 52.5 & 7.53 & 32.2 & 6.88 \\
\ion{Fe}{1} & 6173.34 & 2.22 & $-2.90$ & 70.2 & 7.49 & 53.6 & 6.89 \\
\ion{Fe}{2} & 6149.23 & 3.89 & $-2.80$ & 38.0 & 7.59 & 24.0 & 6.86 \\
\ion{Fe}{2} & 5100.66 & 2.81 & $-4.13$ & 20.3 & 7.47 & 8.7 & 6.61 \\
\ion{Y}{2} & 5087.43 & 1.08 & $-0.36$ & 48.7 & 2.31 & 31.7 & 1.36 \\
\ion{Zr}{2} & 5112.27 & 1.66 & $-0.76$ & 10.0 & 2.64 & 2.3 & 1.49 \\
\ion{Ba}{2} & 6141.73 & 0.70 & $-0.077$ & 127.9 & 2.49 & 94.6 & 1.36 \\ 
\enddata
\label{G265-39 data}
\end{deluxetable}


\begin{thebibliography}{} 

\bibitem[Allen, Poveda \& Herrera(2000)]{Allen2000} Allen, C., Poveda, A., \& Herrera, M. A. 2000, \aap, 356, 529

\bibitem[Alonso \& Martinez-Roger(1996)]{Alonso1996} Alonso, A., \& Martinez-Roger, C. 1996, A\&AS, 117, 227

\bibitem[Arnadottir, Feltzing \& Lundstrom(2010)]{Arnadottir2010} Arnadottir, A. S., Feltzing, S., \& Lundstrom, I. 2010, \aap, 521, 40

\bibitem[Axer, Fuhrmann \& Gehren(1994)]{Axer1994} Axer, M., Fuhrmann, K., \& Gehren, T. 1994, \aap, 291, 895

\bibitem[Beers et al.(1999)]{Beers1999} Beers, T. C., Rossi, S., Norris, J. E., Ryan, S. G., \& Shefler, T. 1999, \aj, 117, 981

\bibitem[Bensby et al.(2005)]{Bensby2005} Bensby, T., Feltzing, S., Lundstroem, I., \& Ilyin, I. 2005, \aap, 433, 185

\bibitem[Boesgaard, Stephens \& Deliyannis(2005)]{Boesgaard2005} Boesgaard, A. M., Stephens, A., \& Deliyannis, C. P. 2005, \apj, 633, 398

\bibitem[Boesgaard(2007)]{Boes2007} Boesgaard, A. M. 2007, \apj, 667, 1196

\bibitem[Boesgaard et al.(2010)]{Boes2010} Boesgaard, A. M., Rich, J. A., Levesque, E. M., \& Bowler, B. P. 2010, IAUS, 268, 231


\bibitem[Caffau et al.(2005)]{Caffau2005} Caffau, E.; Bonifacio, P.; Faraggiana, R.; François, P.; Gratton, R. G.; Barbieri, M. 2005,
\aap, 441, 533

\bibitem[Carlsson et al.(1994)]{Carlsson1994} Carlsson, M., Rutten, R. J., Bruls, J. H. M. J., \& Shchukina, N. G. 1994, \aap, 288, 860

\bibitem[Carney et al.(1994)]{Carney1994} Carney, B. W., Latham, D. W., Laird, J. B., \& Aguilar, L. A. 1994, \aj, 107, 2240

\bibitem[Carney et al.(1997)]{Carney1997} Carney, B. W., Wright, J. S., Sneden, C., Laird, J. B., Aguilar, L. A., \& Latham, D. W. 1997, \aj, 114, 363

\bibitem[Casagrande et al.(2010)]{Casagrande2010} Casagrande, L., Ramirez, I., Melendez, J., Bessell, M., \& Asplund, M. 2010, \aap, 512, 54

\bibitem[Cavallo, Pilachowski \& Rebolo(1997)]{Cavallo97} Cavallo, R. M., Pilachowski, C. A., \& Rebolo, R. 1997, \pasp, 109, 226

\bibitem[Cenarro et al.(2007)]{Cenarro2007} Cenarro, A. J., Peletier, R. F., Sanchez-Blazquez, P., Selam, S. O., Toloba, E., Cardiel, N., Falcon-Barroso, J., Gorgas, J., Jimenez-Vicente, J., \& Vazdekis, A. 2007, \mnras, 374, 664

\bibitem[Clementini et al.(1999)]{Clem1999} Clementini, G., Gratton, R. G., Carretta, E., \& Sneden, C. 1999, \mnras, 302, 22

%\bibitem[D'Antona \& Mazzitelli(1997)]{DM97} D'Antona, F., \& Mazzitelli, I.\ 1997, \memsai, 68, 807 

\bibitem[Deliyannis, Demarque \& Kawaler(1990)]{Deli90} Deliyannis, C. P., Demarque, P., \& Kawaler, S. D. 1990, \apjs, 73, 21

\bibitem[Demarque et al.(2004)]{Demarque2004} Demarque, P., Woo, J.-H., Kim, Y.-C., \& Yi., S. K. 2004, \apjs, 155, 667

\bibitem[Dunkley et al.(2009)]{Dunk2009} Dunkley, J., Komatsu, E., Nolta, M. R., Spergel, D. N., Larson, D., Hinshaw, G., Page, L., Bennett, C. L., et al.~2009, \apjs, 180, 306

\bibitem[Edvardsson et al.(1993)]{Edvard1993} Edvardsson, B., Andersen, J., Gustafsson, B., Lambert, D. L., Nissen, P. E., \& Tomkin, J. 1993, \aap, 275, 101

\bibitem[Fuhrmann(1998)]{Fuhrmann98} Fuhrmann, K. 1998, \aap, 338, 161

\bibitem[Fulbright(2000)]{Fulbright00} Fulbright, J. P. 2000, \aj, 120, 1841

\bibitem[Goldberg et al.(2002)]{Goldberg2002} Goldberg, D., Mazeh, T., Latham, D. W., Stefanik, R. P., Carney, B. W., \& Laird, J. B. 2002, \aj, 124, 1132

\bibitem[Gratton, Carretta \& Castelli(1996)]{Gratton1996} Gratton, R. G., Carretta, E., \& Castelli, F. 1996, \aap, 314, 191

\bibitem[Gratton et al.(2000)]{Gratton2000} Gratton, R. G., Sneden, C., Carretta, E., \& Bragaglia, A. 2000, \aap, 354, 169

\bibitem[Gratton et al.(2003)]{Gratton2003} Gratton, R. G., Carretta, E., Claudi, R., Lucatello, S., \& Barbieri, M. 2003, \aap, 404, 187

\bibitem[Gutierrez et al.(1999)]{Gutierrez1999} Gutierrez, C. M., Garcia Lopez, R. J., Rebolo, R., Martin, E. L., \& Francois, P. 1999, A\&AS, 137, 93

\bibitem[Horch et al.(2002)]{Horch2002} Horch, E. P., Robinson, S. E., Meyer, R. D., van Altena, W. F., Ninkov, Z., \& Piterman, A. 2002, \aj, 123, 3442

\bibitem[Hosford et al.(2009)]{Hosford2009} Hosford, A., Ryan, S. G., Garcia Perez, A. E., Norris, J. E., \& 
Olive, K. A. 2009, \aap, 493, 601

\bibitem[Hosford et al.(2010)]{Hosford2010} Hosford, A., Garcia Perez, A. E., Collet, R., Ryan, S. G., Norris, J. E., \& Olive, K. A. 2010, \aap, 511, 47

\bibitem[Ivans et al.(2003)]{Ivans2003} Ivans, I. I., Sneden, C., James, C. R., Preston, G. W., Fulbright, J. P., Hoeflich, P. A., Carney, B. W., \& Wheeler, J. C. 2003, \apj, 592, 906

\bibitem[Ishigaki, Chiba \& Aoki(2010)]{Ishigaki2010} Ishigaki, M., Chiba, M., \& Aoki, W. 2010, \pasj, 62, 143

\bibitem[James(2000)]{James2000} James, C. R. 2000, Ph. D. dissertation, U. Texas Austin

\bibitem[Jones, Wyse \& Gilmore(1995)]{Jones1995} Jones, J. B., Wyse, R. F. G., \& Gilmore, G. 1995, \pasp, 107, 632

\bibitem[Jonesell et al.(2005)]{Jonsell2005} Jonsell, K., Edvardsson, B., Gustafsson, B., Magain, P., Nissen, P. E.,
\& Asplund, M. 2005, \aap, 440, 321

\bibitem[King et al.(2010)]{KSHP10} King, J. R., Schuler, S. C., Hobbs, L. M., \& Pinsonneault, M. H. 2010, \apj, 710, 1610 

\bibitem[King(1993)]{King93} King, J. R., 1993, \aj, 106, 1206

\bibitem[King(1994)]{King94} King, J. R. 1994, \aj, 107, 1165

\bibitem[King et al.(1996)]{K96} King, J. R., Deliyannis, C. P., \& Boesgaard, A. M. 1996, \aj, 112, 2839

\bibitem[King et al.(1997)]{King1997} King, J. R., Deliyannis, C. P., Hiltgen, D. D., Stephens, A., Cunha, K., \& Boesgaard, A. M. 1997, \aj, 113, 1871

\bibitem[King et al.(1998)]{King1998} King, J. R., Stephens, A., Boesgaard, A. M., \& Deliyannis, C. P. 1998, \aj, 115, 666

\bibitem[Koch, Lind \& Rich(2011)]{Koch2011} Koch, A., Lind, K., \& Rich, R. M. 2011, \apj, 738, L29

\bibitem[Kupka et al.(1999)]{Kupka1999} Kupka, F., Piskunov, N., Ryabchikova, T. A., Stempels, H. C., \& Weiss, W. W. 1999, A\&AS, 138, 119 

\bibitem[Lambert, Heath \& Edvardsson(1991)]{LHE91} Lambert, D. L., Heath, J. E., \& Edvardsson, B. 1991, \mnras, 253, 610

\bibitem[Latham et al.(2002)]{Latham2002} Latham, D. W., Stefanik, R. P., Torres, G., Davis, R. J., Mazeh, T., Carney, B. W., Laird, J. B., \& Morse, J. A. 2002, \aj, 124, 1144

\bibitem[Lejeune et al.(1998)]{Lejeune1998} Lejeune, Th., Cuisinier, F., \& Buser, R. 1998, A\&AS, 130, 65

\bibitem[Mandell, Ge \& Murray(2004)]{Mandell2004} Mandell, A. M., Ge, J., \& Murray, N. 2004, \aj, 127, 1147

\bibitem[McCarthy et al.(1993)]{McCarthy1993} McCarthy, J. K., Sandiford, B. A., Boyd, D., \& Booth, J. 1993, \pasp, 105, 881

\bibitem[Melendez \& Ramirez(2004)]{Melendez2004} Melendez, J., \& Ramirez, I. 2004, \apj, 615, L33

\bibitem[Melendez et al.(2010)]{Melendez2010} Melendez, J., Casagrande, L., Ramirez, I., Asplund, M., \& Schuster, W. J. 2010, \aap, 515, L3

\bibitem[Mishenina et al.(2000)]{Mish2000} Mishenina, T. V., Korotin, S. A., Klochkova, V. G., \& Panchuk, V. E. 2000, \aap, 353, 978

\bibitem[Mishenina \& Kovtyukh(2001)]{Mishenina01} Mishenina, T. V., \& Kovtyukh, V. V. 2001, \aap, 370, 951

\bibitem[Nissen et al.(2004)]{Nissen2004} Nissen, P. E.; Chen, Y. Q.; Asplund, M.; Pettini, M. 2004, \aap, 415, 993

\bibitem[Nissen et al.(2007)]{Nissen2007} Nissen, P. E., Akerman, C., Asplund, M., Fabbian, D., Kerber, F., 
Kaeufl, H. U., \& Pettini, M. 2007, \aap, 469, 319 

\bibitem[Nissen \& Schuster(2010)]{Nissen2010} Nissen, P. E., \& Schuster, W. J. 2010, \aap, 511, 10

\bibitem[Pilachowski, Sneden \& Booth(1993)]{Pilach1993} Pilachowski, C. A., Sneden, C., \& Booth, J. 1993, \apj, 407, 699

\bibitem[Pinsonneault, Deliyannis \& Demarque(1992)]{Pin92} Pinsonneault, M. H., Deliyannis, C. P., \& Demarque, P. 1992, \apjs, 78, 179

\bibitem[Rastegaev, Balega \& Malogolovets(2007)]{Rastegaev2007} Rastegaev, D. A., Balega, Yu. Yu., \& Malogolovets, E. V. 2007, AstBu, 62, 235

\bibitem[Rebolo, Beckman \& Molaro(1988)]{RMB1988} Rebolo, R., Beckman, J. E., \& Molaro, P. 1988, \aap, 192, 192

\bibitem[Ramirez, Allende Prieto \& Lambert, D. L.(2007)]{Ramirez2007} Ramirez, I., Allende Prieto, C., \& Lambert, D. L. 2007, \aap, 465, 271

\bibitem[Reddy, Lambert \& Prieto(2006)]{Reddy2006} Reddy, B. E., Lambert, D. L., \& Prieto, C. A. 2006, \mnras, 367, 1329

\bibitem[Reddy \& Lambert(2008)]{Reddy2008} Reddy, B. E., \& Lambert, D. L. 2008, \mnras, 391, 95

\bibitem[Ryan \& Norris(1991)]{Ryan1991} Ryan, S. G., \& Norris, J. E. 1991, \aj, 101, 1865

\bibitem[Ryan \& Deliyannis(1995)]{RD95} Ryan, S. G., \& Deliyannis, C. P. 1995, \apj, 453, 819 

\bibitem[Ryan et al.(1999)]{Ryan1999} Ryan, S. G., Norris, J. E., Beers, T. C., 1999, \apj, 523, 654

\bibitem[Sbordone et al.(2010)]{Sbordone2010} Sbordone, L, Bonifacio, P., Caffau, E., Ludwig, H.-G., Behara, N. T., Gonzalez Hernando, J. I., Steffen, M., Cayrel, R., et al.~2010, \aap, 522, A26

\bibitem[Schlegel, Finkbeiner, \& Davis(1998)]{Schlegel1998} Schlegel, D. J., Finkbeiner, D. P., \& Davis, M. 1998, \apj, 500, 525

\bibitem[Schuler et al.(2006)]{Sch06} Schuler, S. C., King, J. R., Terndrup, D. M., Pinsonneault, M. H., Murray, N., \& Hobbs, L. M. 2006, \apj, 636, 432

\bibitem[Schuster \& Nissen(1989)]{SN89} Schuster, W. J., \& Nissen, P. E. 1989, \aap, 222, 69

\bibitem[Schuster, Parrao \& Contreras Martinez(1993)]{Schuster1993} Schuster, W. J., Parrao, L., \& Contreras Martinez, M. E. 1993, A\&AS, 97, 951

\bibitem[Schuster et al.(2004)]{Schuster2004} Schuster, W. J., Beers, T. C., Michel, R., Nissen, P. E., \& García, G. 2004, A\&A, 422, 527

\bibitem[Schuster et al.(2006)]{Schuster2006} Schuster, W. J., Moitinho, A., Marquez, A., Parrao, L., \& Covarrubias, E. 2006, \aap, 445, 939

\bibitem[Shi et al.(2007)]{Shi2007} Shi, J. R., Gehren, T., Zhang, H. W., Zeng, J. L., \& Zhao, G. 2007, \aap, 465, 587

\bibitem[Simmerer et al.(2004)]{Simmerer2004} Simmerer, J., Sneden, C., Cowan, J. J., Collier, J., Woolf, V. M., \& Lawler, J. E. 2004, \apj, 617, 1091

\bibitem[Smiljanic et al.(2009)]{Smil2009} Smiljanic, R., Pasquini, L., Bonifacio, P., Galli, D., Gratton, R. G., Randich, S., 
\& Wolff, B. 2009, \aap, 499, 103

\bibitem[Sneden(1973)]{Sneden1973} Sneden, C. 1973, \apj, 184, 839

\bibitem[Sousa et al.(2011)]{Sousa2011} Sousa, S. G., Santos, N. C., Israelian, G., Lovis, C., Mayor, M., Silva, P. B., \& 
Udry, S. 2005, \aap, 526, 99

\bibitem[Spite \& Spite(1982)]{SS82} Spite, M., \& Spite, F. 1982, Nature, 297, 483

\bibitem[Stephens et al.(1997)]{Stephens97} Stephens, A., Boesgaard, A. M., King, J. R., \& Deliyannis, C. P. 1997, \apj, 491, 339

\bibitem[Stephens \& Boesgaard(2002)]{Stephens2002} Stephens, A., \& Boesgaard, A. M. 2002, \aj, 123, 1647

\bibitem[Tomkin et al.(1992)]{Tomkin92} Tomkin, J., Lemke, M., Lambert, D. L., \& Sneden, C. 1992, \aj, 104, 1568

\bibitem[Tull et al.(1995)]{Tull1995} Tull, R. G., MacQueen, P. J., Sneden, C., \& Lambert, D. L. 1995, \pasp, 107, 251

\bibitem[Valenti \& Fischer(2005)]{VF2005} Valenti, J. A., \& Fischer, D. A. 2005, \apjs, 159, 141

\bibitem[White, Gabor \& Hillenbrand(2007)]{White2007} White, R. J., Gabor, J. M., \& Hillenbrand, L. A. 2007, aj, 133, 2524

\bibitem[Yong \& Lambert(2003)]{Yong2003} Yong, D., \& Lambert, D. L. 2003, \pasp, 115, 22

\bibitem[Zhang et al.(2009)]{Zhang2009} Zhang, L., Ishigaki, M., Aoki, W., Zhao, G., \& Chiba, M. 2009, \apj, 706, 1095

\bibitem[Zhang \& Zhao(2006)]{Zhang2006} Zhang, H. W., \& Zhao, G. 2006, \aap, 449, 127

\bibitem[Zhang \& Zhao(2005)]{Zhang2005} Zhang, H. W., \& Zhao, G. 2005, \mnras, 364, 712

\bibitem[Zhang \& Zhao(2003)]{Zhang2003} Zhang, H.-W, \& Zhao, G. 2003, ChJAA, 3, 453

\end{thebibliography}
\end{document}